\documentclass[final,3p,times,numbers,sort,compress]{elsarticle}

\usepackage{amsmath}
\usepackage{amsfonts}
\usepackage{amsmath}
\usepackage{amssymb}
\usepackage{amsthm}
\usepackage{dcolumn}
\usepackage{bm}
\usepackage{comment}
\usepackage{graphicx}
\usepackage[breaklinks=true,colorlinks=true,linkcolor=blue,urlcolor=blue,citecolor=blue]{hyperref}
\usepackage{ulem}

\newcommand{\defeq}{\mathrel{\mathop:}=}

\begin{document}

\begin{frontmatter}

\title{Microcanonical Monte Carlo simulation of\\ opinion dynamics under the influence of mass media}

\author[ificc]{Yasmín Navarrete\corref{cor1}}
\ead{yasmin.navarrete@gmail.com}

\address[ificc]{Instituto de Filosofía y Ciencias de la Complejidad, Los Alerces 3024, Ñuñoa, Santiago, Chile.}
\address[unab]{Departamento de F\'isica, Facultad de Ciencias Exactas, Universidad Andres Bello,\\ Sazi\'e 2212, piso 7, 8370136, Santiago, Chile.}
\address[cchen]{Research Center on the Intersection in Plasma Physics, Matter and Complexity, P$^2$mc,\\ Comisión Chilena de Energía Nuclear, Casilla 188-D, Santiago, Chile}

\author[unab]{Carlos Femenías}
\author[cchen,unab]{Sergio Davis}
\author[unab]{Claudia Loyola}

\begin{abstract}
The formation of large social groups having uniform opinions influenced by mass media is currently an important topic in the social sciences.
In this work, we explore and extend an off-lattice, two-dimensional Potts model (Eur. Phys. J. B \textbf{87}, 78 [2014]) that describes the formation and dynamics 
of opinions in social groups according to individual consequence and agreement between neighbors. This model was originally obtained by the application of the 
maximum entropy principle, a general method in statistical inference, and using the same methodology we have now included the influence of mass media as a constant 
external field. By means of microcanonical Monte Carlo Metropolis simulations on a setup with two regions with opposing external influences, we have shown the presence 
of metastable states associated to the formation of clusters aligned with the locally imposed opinion. Our results suggest that, for some values of the total energy 
of the system, only a single cluster with a uniform opinion survives, thus the presence of two large, opposing groups is not a thermodynamically stable configuration.
\end{abstract}

\end{frontmatter}



%
%
%
\section{Introduction}
\label{sec:introduction}

In the study of social models, it has become increasingly common to apply physical frameworks to gain a deeper understanding of social phenomena, a field known as sociophysics~\cite{Schewitzer2018}. This approach has proven valuable in addressing fundamental questions about the formation, stability, and evolution of social groups driven by individual interactions. Here, we propose integrating external influences, such as media, social platforms, or environmental factors, as an external field to analyze their impact on social structure dynamics. This extension offers a more detailed view of how external factors influence social behavior and group dynamics.

Modern opinion dynamics models, such as the Sznajd and Axelrod models, are essential for understanding how opinions and cultural traits spread~\cite{Sznajd2000, Axelrod1997}. These models have been widely studied and have been adapted to various networks, incorporating factors like reputation and independence, social 
temperature, anti-conformity, and cultural drift, providing deeper insights into consensus, disagreement, and cultural evolution~\cite{Crokidakis2012, Sirbu2016, 
Pineda2015, Evans2018}. Various numerical approaches have been applied to these models, including non-deterministic cellular automata~\cite{Kacperski1999}, variants 
of the Potts model~\cite{Liu2001,Schulze2005}, and the Langevin equation for describing stochastic dynamics~\cite{Hu2009}. In addition, Hamiltonian models
~\cite{Fronczak2006} and other statistical mechanics tools, along with different empirical models analogous to physical phenomena~\cite{Ausloos2007,Ausloos2009,Davis2014}, have been utilized. These models exhibit numerous fascinating emergent properties. They reveal various phase transitions, offering a deeper understanding 
of phenomena like state ordering and their evolution toward equilibrium. 

However, recent research considering social mobility has changed from fixed-lattice models to adaptive networks. As new elements are incorporated, these models remain vital and 
significant in advancing our comprehension of complex systems~\cite{Sirbu2016, Pinto2016}. 
These new approaches allow for a more nuanced representation of how individuals and groups navigate social interactions, taking into account the mass media as an external permanent magnetic field in analogy. These models provide a deeper understanding of an agent's mobility mechanisms by incorporating elements such as variable social ties, peer influence, and the impact of external factors potentially being economic changes and technological advancements. As these adaptive models continue to develop, they remain significant in advancing our comprehension of social systems, offering valuable insights into the intricate interplay between individual agency and structural constraints. This progress not only improves our theoretical knowledge but also has practical implications in different fields such as social science\cite{NICOLAO2019, Liu2001}, biology and other sciences\cite{Utkir2022}.

In this work, we aim to extend a previously developed social model~\cite{Davis2014, Farias2021} that considers individual influence, the significance of opinions, and the interactions between individuals, incorporating external influences such as social media~\cite{Pinto2016}. To achieve this, we again used the statistical mechanics approach to create 
a Hamiltonian objective function to describe equilibrium states, spatial distributions, and the evolution of order and disorder in a society. Our approach includes 
expanding the Potts model~\cite{Wu1982} to incorporate agent mobility and internal resistance to neighbor influence, aiming for a fundamental approach to modeling social 
dynamics, together with incorporating an external field representing social media, namely external influence. In the subsequent sections, we will elaborate on our 
methodology in Section~\ref{sec:methodology}, present and discuss our results in Section~\ref{sec:results} and closing with some implications and insights from our 
findings in Section~\ref{sec:conclusions}.

\section{Model and Method}
\label{sec:methodology}

\nocite{Davis2014}
\nocite{Farias2021}

In our setup, we have a collection of $N$ agents (or individuals), each able to move freely within a two-dimensional box of length $L$. These agents have the freedom to move freely inside the box, unlike those constrained by a lattice structure. Each agent possesses several attributes: their position ($\mathbf{r}_i$), personal belief ($B_i$), external behavior ($S_i$), consequence level ($C_i$), and social personality ($J_i$). In addition, the box is divided into the left and right halves such that
there are external permanent influences defined on each half, particularly one acting on the side $[0, L/2]$ with intensity $G_{\text{left}}$ and the other on the
side $[L/2, L]$ with intensity $G_{\text{right}}$.

Both personal belief $B_i$ and external behavior $S_i$ are discrete, taking values between 1 and $Q$, where $Q$ represents the number of potential beliefs. The consequence level $C_i$ assumes non-negative continuous values, reflecting how the external behavior of an agent aligns with personal beliefs; it quantifies the energy cost associated with any discrepancy. The social personality $J_i$ manifests itself as continuous values; Positive values prompt agents to gravitate toward others with similar external behaviors, while negative values attract them toward those with different external behaviors.

Within the model, the following assumptions are made:

\begin{enumerate}
\item External behavior does not necessarily align with personal beliefs, and the cost of internal disagreement is directly proportional to $C$, highlighting the tension between outward conformity and inner conviction.
\item Agents beyond a critical distance $R_c$ cannot influence each other, emphasizing the significance of proximity in social interactions and ensuring localized effects.
\item The coupling parameter $J>0$ quantifies the cost of disagreeing with other agents, underscoring the importance of consensus and social harmony within the group.
\item Agents have the potential to align their opinions with a persistent external influence, demonstrating the impact of constant external pressures on individual belief systems.
\item Agents possess the freedom to move without the constraints of an underlying lattice, allowing for dynamic interactions and a more realistic representation of social mobility.
\end{enumerate}

\noindent
These five postulates motivate the following energy function or Hamiltonian for a system of $N$ agents,
\begin{equation}
\begin{aligned}
\mathcal{H} = & -\frac{1}{2}\sum_{i=1}^N \sum_{j\neq i}
J \Theta(R_c-\vert\mathbf{r}_j-\mathbf{r}_i\vert)\left<S_i S_j\right> \\
& - \sum_{i=1}^N \Bigg[C\left<S_i, B_i\right>
- G_{\text{left}}\Theta\left(\frac{L}{2}-x_i\right)
- G_{\text{right}}\Theta\left(x_i-\frac{L}{2}\right)\Bigg]
\end{aligned}
\end{equation}
where we have assumed homogeneity among agents, that is, that $C_i = C$ and $J_i = J$ for all $i$. For more details on obtaining the Hamiltonian, see \ref{Appendix}. Here we have defined the 
product
\begin{equation}
\left<a, b\right> \defeq 2\delta(a, b)-1,
\end{equation}
with $\delta$ the Kronecker delta, $\Theta$ is the Heaviside step function and $S_{\text{left}}$ and $S_{\text{right}}$ are the opinions imposed by the external influences on the left and right sides of the box, respectively. The last two terms are newly introduced in this work. In order to understand the thermodynamics of the system, we start with the Gibbs-Shannon entropy, defined by
\begin{equation}
S = - \sum_{\chi} P(\chi) \ln P(\chi)
\end{equation}
where $\chi$ represents the microscopic state combining the discrete opinion variables and the continuous position variables, and maximize it under the constraints mentioned above~\cite{Jaynes1957}. We find that 
the only independent thermodynamic variables are $N$, $Q$, $\rho = R_c/L$, $\gamma = J/C$, and $\tau = T/C$. This allows us to define a \textit{social temperature} and entropy, which plays a fundamental role in the thermodynamic characterization of the observed phenomena. 

In our earlier research~\cite{Davis2014, Farias2021}, we observed and documented how thermodynamic parameters are interconnected with different social phenomena. 
This information is summarized in Table~\ref{tbl:params} below for easy reference.

\begin{table}[h!]
\centering

\begin{tabular}{c|c|c}
\hline
\textbf{Symbol} & \textbf{Social meaning} & \textbf{Thermodynamic analog} \\
\hline\hline
$\bm{r}_i$ & Position & Particle position \\
$S_i$ & External opinion & Particle spin orientation \\
$B_i$ & Internal opinion & Particle local field \\
\hline
$N$ & Number of agents & Number of particles \\
$Q$ & Number of possible opinions & Magnitude of spin \\
$R_c$ & Influence radius & Interaction cutoff \\
$C$ & Consequence level & Local anisotropy \\
$J$ & Social personality & Exchange coupling \\
$T$ & Social temperature & Temperature \\
$E$ & Dissatisfaction & Internal energy \\
$G$ & Mass media influence & External magnetic field\\
\hline
\end{tabular}
\caption{Parameters in the model, their meaning and thermodynamic analogs.}
\label{tbl:params}
\end{table}

\section{Results}
\label{sec:results}

\begin{figure}[h!]
    \centering
    \includegraphics[width=0.6\textwidth]{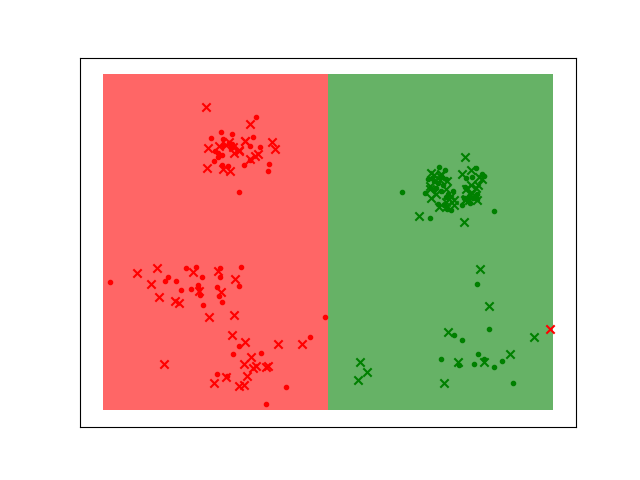}
    \caption{Initial configuration of the system with a preparation energy of -8000 (arbitrary units), with a total energy of -5000 (arbitrary units). The red and green regions represent the mass media influence, and the dots and crosses represent the agents in the regions.}
    \label{fig:initial}
\end{figure}

Using microcanonical Monte Carlo Metropolis simulations, we examined a setup with two regions, each subjected to opposing external influences representative of mass media. The figure~\ref{fig:initial} represents an initial structure with two delimited areas, red and green, related to the $G_\text{left}$ and $G_\text{right}$ mass media influence, respectively. A dot, a cross, and a specific color represent the agents in the regions. An agent of red or green color represents the public opinion (red or green), and the cross represents a public opinion different from the internal opinion, which can be interpreted as a fake agent. Finally, the dot shows a sincere agent, in which the public opinion is the same as the internal one. The initial system is prepared by minimizing the agents energy on the region, to obtain a minimum energy configuration. The structure is prepared with a total energy of -8000 (arbitrary units) and a total of -5000 (arbitrary units).

In all cases, the rejection rate of the Metropolis algorithm was close to 50\%, providing an adequate sampling of states in the neighborhood around the most probable state.

\begin{figure}[h!]
    \centering
    \begin{minipage}[t]{0.47\textwidth}
        \centering
        \includegraphics[width=\textwidth]{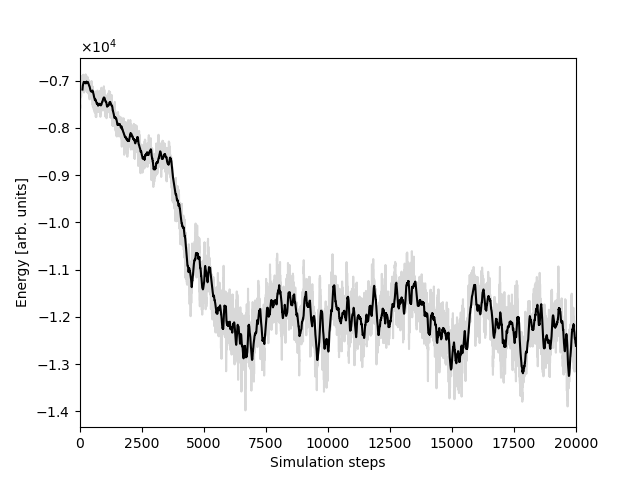}
        \caption{Energy of the system, the black thick line corresponds to a mobile average of 100 steps for clarity in the energy. As we observe a steepest }
        \label{fig:energy_plot}
    \end{minipage}%
    \hspace{0.04\textwidth} 
    \begin{minipage}[t]{0.47\textwidth}
        \centering
        \includegraphics[width=\textwidth]{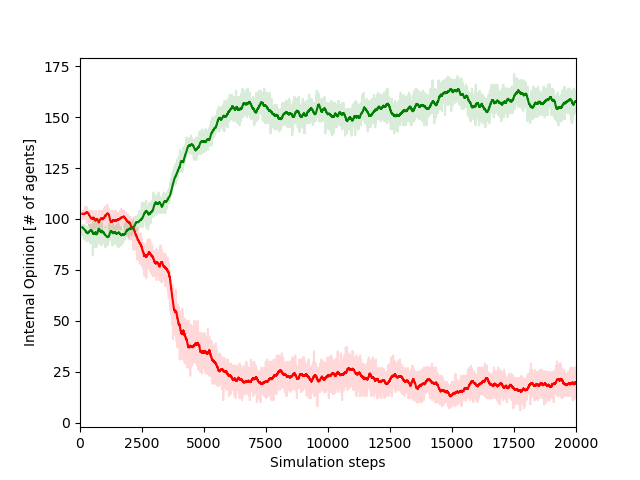}
        \caption{Internal opinion of agents as the simulation progresses. The green and red lines represent the agents with two different opinions. For clarity, the thick green and red lines correspond to a moving average over 100 steps.}
        \label{fig:b_plot}
    \end{minipage}\\[2ex]
\end{figure}

Figures \ref{fig:energy_plot}, and \ref{fig:b_plot} present the evolution of information about the system throughout the simulation. In Figure \ref{fig:energy_plot}, we observe how the potential energy of the system decreases progressively as the simulation evolves, which is expected during a minimization process. As the simulation progresses, the energy stabilizes, indicating that the system is approaching a more stable state.

Figure \ref{fig:b_plot} shows the internal opinion dynamics of the agents. The green line represents the opinion of one group of agents, while the red line corresponds to another. The agents tend to align with one of the major opinions (green region in this case). However, some agents remain in their initial states, demonstrating the presence of individuals with strong convictions who resist shifting to the dominant opinion.

\begin{figure}
    \centering
    \includegraphics[width=0.6\linewidth]{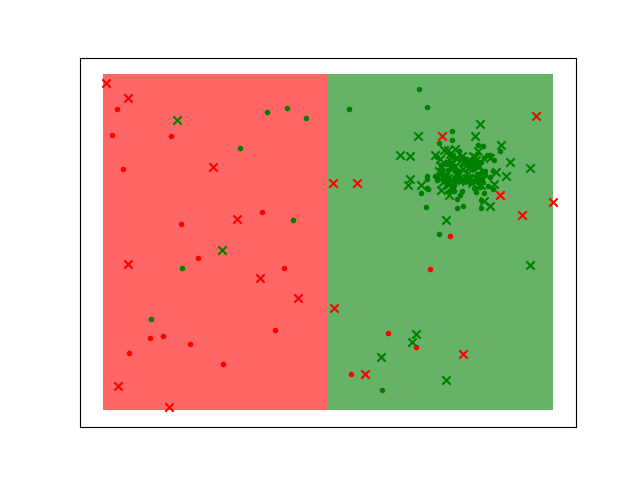}
    \caption{ Final configuration of the system showing the distribution of agents after the influence process. The red and green regions still represent the mass media influence, but the agents (dots and crosses) have redistributed, indicating the influence dynamics over time. The clusters and dispersion of agents highlight the change in opinion influenced by the contrasting regions.}
    \label{fig:final}
\end{figure}

Additionally to the relevant information that could be extracted from the simulations, a final configuration of the final state can also be observed at figure~\ref{fig:final}. Here, we observe how most agents are displaced to the green area, building a solid cluster of agents with a firm personal conviction (B value?). As we observe in this configuration, the dot green points are mainly surrounded by cross-green agents, which makes it difficult to displace red agents that can modify or destroy the group.

We observed metastable states characterized by the formation of clusters with opinions aligned to the local external influence. This indicates that clusters of uniform opinion can persist over time, even in the presence of opposing influences. This could be seen as a form of \textit{social conformity}, where individuals lose their personal beliefs, internal identity, and unique relationships, ultimately aligning with a dominant opinion. Note that for certain values of the total energy of the system, our simulations revealed that only a single cluster with a uniform opinion survives. This outcome suggests that the configuration with two large, opposing opinion groups is not thermodynamically stable.
As a matter of fact, the stability of opinion clusters is heavily dependent on the total energy of the system. The presence of a single uniform opinion cluster occurs within specific energy ranges.

A minimum level of consistency $C$ is required for individuals to change their opinion. Below this level, people either follow the crowd or remain independent, but never both. However, when $C$ exceeds this critical value, continuous phase transitions occur, where behavior is governed by beliefs instead of 
social pressure, allowing individual opinions to persist. This persistence of opinions in high social consistency leads us to explore opinion dynamics within social groups influenced by mass media. Including mass media as an external field significantly impacts the behavior and stability of opinion clusters. Our results show metastable states -characterized by clusters of agents- sharing a uniform opinion aligned with the local external influence, as is naturally expected; however, clusters with a different opinion that is not aligned can also be observed in the system. 

The configuration with two large opposing groups is not thermodynamically stable, and only a single, homogeneous group survives. This can be explained by a mechanism in 
which the boundary between the two influence zones acts as a membrane through which individual agents can pass and change their opinion, leading to the gradual vanishing 
of one of the groups. 

\section{Conclusions}
\label{sec:conclusions}

We have analyzed the model initially proposed in Ref.~\cite{Davis2014} and whose metastable states were studied in Ref.~\cite{Farias2021}, where external influences due to mass media are included. We have found that the configuration with two large opposing groups is not thermodynamically stable and that only a single, homogeneous group survives. We have explained this through a mechanism in which the boundary between the two influence zones acts as a membrane through which individual agents can pass through and change their opinion, leading to one of the groups slowly vanishing.
Our results indicate that mass media can significantly influence the formation and stability of opinion clusters within social groups. The introduction of mass media as an external field can lead to the dominance of a single opinion group, reducing the likelihood of stable coexistence of opposing opinions. This has important implications for understanding the impact of mass media on social cohesion and the dynamics of opinion formation.

By incorporating the influence of mass media as an external field, our model provides a more comprehensive framework for understanding the dynamics of opinion formation under external influences.

\section*{Acknowledgments}

\noindent
The authors acknowledge financial support from ANID FONDECYT 1220651 grant. Computational work was supported by the supercomputing infrastructures of the NLHPC (ECM-02) and FENIX (UNAB).

\appendix
\section{Derivation of the external field interaction terms in the Hamiltonian}\label{Appendix}

In this section, we develop the mathematical formulation to find the Hamiltonian of the system, particularly the expected values for the external field interaction. 
As is depicted in Ref. \cite{Davis2014}, we have the following Hamiltonian, when the external field interaction is not considered.

\begin{equation}
\bar{\mathcal{H}} = -\frac{1}{2} J \sum_{i=1}^{N} \sum_{\substack{j \neq i}} \left< S_i, S_j \right> 
+ \frac{1}{2} \sum_{i=1}^{N} \sum_{\substack{j \neq i}} R 
- C \sum_{i=1}^{N} \left< S_i, B_i \right>.
\end{equation}
where the three parameters $J$, $C$ and $R$ are defined in terms of the original Lagrange multipliers by
\begin{subequations}
\begin{align}
J & \defeq \frac{\lambda_J}{\beta N (N - 1)}, \\
C & \defeq \frac{\lambda_C}{2 \beta N}, \\
R & \defeq \frac{- \lambda_J + 2 \lambda_R}{\beta N (N - 1)}.
\end{align}
\end{subequations}

\noindent
Now by taking into account the permanent external magnetic field, we have:

\begin{equation}
\begin{aligned}
\mathcal{H}_{\text ext} =
& - \sum_{i=1}^N \Bigg[
- G_{\text{left}}\delta(S_i, S_{\text{left} })\Theta\left(\frac{L}{2}-x_i\right)
- G_{\text{right}}\delta\left (S_i, S_{\text{right}}\right )\Theta\left(x_i-\frac{L}{2}\right)\Bigg],
\end{aligned}
\end{equation}
then for the permanent external magnetic field at the left side we have
\begin{equation}
P \left (S_i = S_{\text{left}} \mid x_i <\frac{L}{2} \right) =
\frac{\left\langle \delta(S_i, S_{\text{left} }) \Theta\left (\frac{L}{2} - x_{i}\right) \right\rangle}
{\left\langle \Theta\left (\frac{L}{2} -x_i \right) \right\rangle}
= P_{\text{left}}.
\end{equation}

\noindent
Hence, the constraint written above can be expressed as
\begin{equation}
\left\langle  \Theta\left (\frac{L}{2} - x_{i}\right) \left ( \delta\left (S_i, S_{\text{left}}\right ) - P_{\text{left}}. \right ) \right\rangle = 0.\label{eq2}
\end{equation}

\noindent
Analogously, for the permanent magnetic field at the right side, we can obtain:
\begin{equation}
\left\langle  \Theta\left (x_{i} -\frac{L}{2} \right) \left ( \delta\left (S_i, S_{\text{right}}\right ) - P_{\text{right}}. \right ) \right\rangle = 0.
\end{equation}

\noindent
Consider the following arrangement,
\begin{equation}
\langle S_i, S_{\text{leftt}}\rangle = 2\delta\left (S_i, S_{\text{leftt}}\right )-1\label{eq1},
\end{equation}
therefore, by using Eq. \ref{eq1} in Eq. \ref{eq2} we have
\begin{equation}
G_{\text{left}}=\frac{1}{2}\left(2 P_{\text{leftt}}-1\right).
\end{equation}

\noindent
By operating in the same way for the right magnetic field, we obtain
\begin{equation}
G_{\text{right}}=\frac{1}{2}\left(2 P_{\text{right}}-1\right),
\end{equation}
then the Hamiltonian for the external magnetic field is given by
\begin{equation}
\begin{aligned}
\mathcal{H}_{\text ext} =
& - \sum_{i=1}^N \Bigg[
- G_{\text{left}}\Theta\left(\frac{L}{2}-x_i\right)
- G_{\text{right}}\Theta\left(x_i-\frac{L}{2}\right)\Bigg].
\end{aligned}
\end{equation}

\noindent
So, by considering the external field and to avoid the overcrowding: between agents, we have $R=0$, then
\begin{equation}
\mathcal{H}=\bar{\mathcal{H}}+\mathcal{H}_{\text{ext}},
\end{equation}
finally obtaining the full Hamiltonian,
\begin{equation}
\begin{split}
\mathcal{H} & = -\frac{1}{2} J \sum_{i=1}^{N} \sum_{\substack{j \neq i}} \left< S_i, S_j \right> 
+ \frac{1}{2} \sum_{i=1}^{N} \sum_{\substack{j \neq i}} R 
- C\sum_{i=1}^N \left< S_i, B_i \right> + \\
&   \sum_{i=1}^N \Bigg[
 G_{\text{left}}\Theta\left(\frac{L}{2}-x_i\right)
+ G_{\text{right}}\Theta\left(x_i-\frac{L}{2}\right)\Bigg].
\end{split}
\end{equation}

%
%
%
\bibliography{influences}

\begin{thebibliography}{21}
\expandafter\ifx\csname natexlab\endcsname\relax\def\natexlab#1{#1}\fi
\expandafter\ifx\csname bibnamefont\endcsname\relax
  \def\bibnamefont#1{#1}\fi
\expandafter\ifx\csname bibfnamefont\endcsname\relax
  \def\bibfnamefont#1{#1}\fi
\expandafter\ifx\csname citenamefont\endcsname\relax
  \def\citenamefont#1{#1}\fi
\expandafter\ifx\csname url\endcsname\relax
  \def\url#1{\texttt{#1}}\fi
\expandafter\ifx\csname urlprefix\endcsname\relax\def\urlprefix{URL }\fi
\providecommand{\bibinfo}[2]{#2}
\providecommand{\eprint}[2][]{\url{#2}}

\bibitem[{\citenamefont{Schweitzer}(2018)}]{Schewitzer2018}
\bibinfo{author}{\bibfnamefont{F.}~\bibnamefont{Schweitzer}},
  \bibinfo{journal}{Physics Today} \textbf{\bibinfo{volume}{71}},
  \bibinfo{pages}{40} (\bibinfo{year}{2018}).

\bibitem[{\citenamefont{Sznajd-Weron and Sznajd}(2000)}]{Sznajd2000}
\bibinfo{author}{\bibfnamefont{K.}~\bibnamefont{Sznajd-Weron}}
  \bibnamefont{and} \bibinfo{author}{\bibfnamefont{J.}~\bibnamefont{Sznajd}},
  \bibinfo{journal}{Int. J. Mod. Phys. C} \textbf{\bibinfo{volume}{11}},
  \bibinfo{pages}{1157} (\bibinfo{year}{2000}).

\bibitem[{\citenamefont{Axelrod}(1997)}]{Axelrod1997}
\bibinfo{author}{\bibfnamefont{R.}~\bibnamefont{Axelrod}}, \bibinfo{journal}{J.
  Conflict Resolut.} \textbf{\bibinfo{volume}{41}}, \bibinfo{pages}{203}
  (\bibinfo{year}{1997}).

\bibitem[{\citenamefont{Crokidakis and Anteneodo}(2012)}]{Crokidakis2012}
\bibinfo{author}{\bibfnamefont{N.}~\bibnamefont{Crokidakis}} \bibnamefont{and}
  \bibinfo{author}{\bibfnamefont{C.}~\bibnamefont{Anteneodo}},
  \bibinfo{journal}{Phys. Rev. E} \textbf{\bibinfo{volume}{86}}
  (\bibinfo{year}{2012}).

\bibitem[{\citenamefont{Sîrbu et~al.}(2016)\citenamefont{Sîrbu, Loreto,
  Servedio, and Tria}}]{Sirbu2016}
\bibinfo{author}{\bibfnamefont{A.}~\bibnamefont{Sîrbu}},
  \bibinfo{author}{\bibfnamefont{V.}~\bibnamefont{Loreto}},
  \bibinfo{author}{\bibfnamefont{V.~D.~P.} \bibnamefont{Servedio}},
  \bibnamefont{and} \bibinfo{author}{\bibfnamefont{F.}~\bibnamefont{Tria}},
  \emph{\bibinfo{title}{Opinion Dynamics: Models, Extensions and External
  Effects}} (\bibinfo{publisher}{Springer International Publishing},
  \bibinfo{year}{2016}), p. \bibinfo{pages}{363–401}.

\bibitem[{\citenamefont{Pineda and Buendía}(2015)}]{Pineda2015}
\bibinfo{author}{\bibfnamefont{M.}~\bibnamefont{Pineda}} \bibnamefont{and}
  \bibinfo{author}{\bibfnamefont{G.~M.} \bibnamefont{Buendía}},
  \bibinfo{journal}{Phys. A} \textbf{\bibinfo{volume}{420}},
  \bibinfo{pages}{73–84} (\bibinfo{year}{2015}).

\bibitem[{\citenamefont{Evans and Fu}(2018)}]{Evans2018}
\bibinfo{author}{\bibfnamefont{T.}~\bibnamefont{Evans}} \bibnamefont{and}
  \bibinfo{author}{\bibfnamefont{F.}~\bibnamefont{Fu}}, \bibinfo{journal}{Royal
  Society Open Science} \textbf{\bibinfo{volume}{5}}, \bibinfo{pages}{181122}
  (\bibinfo{year}{2018}).

\bibitem[{\citenamefont{Kacperski and Holyst}(1999)}]{Kacperski1999}
\bibinfo{author}{\bibfnamefont{K.}~\bibnamefont{Kacperski}} \bibnamefont{and}
  \bibinfo{author}{\bibfnamefont{J.~A.} \bibnamefont{Holyst}},
  \bibinfo{journal}{Physica A} \textbf{\bibinfo{volume}{269}},
  \bibinfo{pages}{511} (\bibinfo{year}{1999}).

\bibitem[{\citenamefont{Liu et~al.}(2001)\citenamefont{Liu, Luo, and
  Shao}}]{Liu2001}
\bibinfo{author}{\bibfnamefont{Z.}~\bibnamefont{Liu}},
  \bibinfo{author}{\bibfnamefont{J.}~\bibnamefont{Luo}}, \bibnamefont{and}
  \bibinfo{author}{\bibfnamefont{C.}~\bibnamefont{Shao}},
  \bibinfo{journal}{Phys. Rev. E} \textbf{\bibinfo{volume}{64}},
  \bibinfo{pages}{046134} (\bibinfo{year}{2001}).

\bibitem[{\citenamefont{Schulze}(2005)}]{Schulze2005}
\bibinfo{author}{\bibfnamefont{C.}~\bibnamefont{Schulze}},
  \bibinfo{journal}{Int. J. Mod. Phys. C} \textbf{\bibinfo{volume}{16}},
  \bibinfo{pages}{351} (\bibinfo{year}{2005}).

\bibitem[{\citenamefont{Hu and Wang}(2009)}]{Hu2009}
\bibinfo{author}{\bibfnamefont{H.-B.} \bibnamefont{Hu}} \bibnamefont{and}
  \bibinfo{author}{\bibfnamefont{X.-F.} \bibnamefont{Wang}},
  \bibinfo{journal}{J. Phys. A: Math. Theor.} \textbf{\bibinfo{volume}{42}},
  \bibinfo{pages}{225005} (\bibinfo{year}{2009}).

\bibitem[{\citenamefont{Fronczak et~al.}(2006)\citenamefont{Fronczak, Fronczak,
  and Holyst}}]{Fronczak2006}
\bibinfo{author}{\bibfnamefont{P.}~\bibnamefont{Fronczak}},
  \bibinfo{author}{\bibfnamefont{A.}~\bibnamefont{Fronczak}}, \bibnamefont{and}
  \bibinfo{author}{\bibfnamefont{J.~A.} \bibnamefont{Holyst}},
  \bibinfo{journal}{Int. J. Mod. Phys. C} \textbf{\bibinfo{volume}{17}},
  \bibinfo{pages}{1227} (\bibinfo{year}{2006}).

\bibitem[{\citenamefont{Ausloos and Petroni}(2007)}]{Ausloos2007}
\bibinfo{author}{\bibfnamefont{M.}~\bibnamefont{Ausloos}} \bibnamefont{and}
  \bibinfo{author}{\bibfnamefont{F.}~\bibnamefont{Petroni}},
  \bibinfo{journal}{EPL} \textbf{\bibinfo{volume}{77}}, \bibinfo{pages}{38002}
  (\bibinfo{year}{2007}).

\bibitem[{\citenamefont{Ausloos and Petroni}(2009)}]{Ausloos2009}
\bibinfo{author}{\bibfnamefont{M.}~\bibnamefont{Ausloos}} \bibnamefont{and}
  \bibinfo{author}{\bibfnamefont{F.}~\bibnamefont{Petroni}},
  \bibinfo{journal}{Physica A} \textbf{\bibinfo{volume}{388}},
  \bibinfo{pages}{4438} (\bibinfo{year}{2009}).

\bibitem[{\citenamefont{Davis et~al.}(2014)\citenamefont{Davis, Navarrete, and
  Gutiérrez}}]{Davis2014}
\bibinfo{author}{\bibfnamefont{S.}~\bibnamefont{Davis}},
  \bibinfo{author}{\bibfnamefont{Y.}~\bibnamefont{Navarrete}},
  \bibnamefont{and}
  \bibinfo{author}{\bibfnamefont{G.}~\bibnamefont{Gutiérrez}},
  \bibinfo{journal}{Eur. Phys. J. B} \textbf{\bibinfo{volume}{87}},
  \bibinfo{pages}{78} (\bibinfo{year}{2014}).

\bibitem[{\citenamefont{Pinto et~al.}(2016)\citenamefont{Pinto, Balenzuela, and
  Dorso}}]{Pinto2016}
\bibinfo{author}{\bibfnamefont{S.}~\bibnamefont{Pinto}},
  \bibinfo{author}{\bibfnamefont{P.}~\bibnamefont{Balenzuela}},
  \bibnamefont{and} \bibinfo{author}{\bibfnamefont{C.~O.} \bibnamefont{Dorso}},
  \bibinfo{journal}{Phys. A} \textbf{\bibinfo{volume}{458}},
  \bibinfo{pages}{378–390} (\bibinfo{year}{2016}).

\bibitem[{\citenamefont{Nicolao and Ostilli}(2019)}]{NICOLAO2019}
\bibinfo{author}{\bibfnamefont{L.}~\bibnamefont{Nicolao}} \bibnamefont{and}
  \bibinfo{author}{\bibfnamefont{M.}~\bibnamefont{Ostilli}},
  \bibinfo{journal}{Phys. A} \textbf{\bibinfo{volume}{533}},
  \bibinfo{pages}{121920} (\bibinfo{year}{2019}).

\bibitem[{\citenamefont{Rozikov}(2022)}]{Utkir2022}
\bibinfo{author}{\bibfnamefont{U.~A.} \bibnamefont{Rozikov}},
  \emph{\bibinfo{title}{Gibbs measures and Potts model}}
  (\bibinfo{year}{2022}), chap.~\bibinfo{chapter}{2}, pp.
  \bibinfo{pages}{15--79}.

\bibitem[{\citenamefont{Farías and Davis}(2021)}]{Farias2021}
\bibinfo{author}{\bibfnamefont{C.}~\bibnamefont{Farías}} \bibnamefont{and}
  \bibinfo{author}{\bibfnamefont{S.}~\bibnamefont{Davis}},
  \bibinfo{journal}{Phys. A} \textbf{\bibinfo{volume}{581}},
  \bibinfo{pages}{126215} (\bibinfo{year}{2021}).

\bibitem[{\citenamefont{Wu}(1982)}]{Wu1982}
\bibinfo{author}{\bibfnamefont{F.~Y.} \bibnamefont{Wu}}, \bibinfo{journal}{Rev.
  Mod. Phys.} \textbf{\bibinfo{volume}{54}}, \bibinfo{pages}{235}
  (\bibinfo{year}{1982}).

\bibitem[{\citenamefont{Jaynes}(1957)}]{Jaynes1957}
\bibinfo{author}{\bibfnamefont{E.~T.} \bibnamefont{Jaynes}},
  \bibinfo{journal}{Physical Review} \textbf{\bibinfo{volume}{106}},
  \bibinfo{pages}{620} (\bibinfo{year}{1957}).

\end{thebibliography}
\bibliographystyle{apsrev}

\end{document}